\documentstyle[12pt,boxedminipage,epsfig,wrapfig,floatflt,shadow]{article}

\begin{document}

\title{Helping students learn effective problem solving strategies by reflecting with peers}
\author{Andrew Mason and Chandralekha Singh\\ Department of Physics and Astronomy\\ University of Pittsburgh, Pittsburgh, PA, 15260}
\date{ }

\maketitle

\vspace*{-.2in}

\begin{abstract}
We describe a study in which introductory physics students engage in reflection with peers about problem solving.
The recitations for an introductory physics course with 200 students were broken into the ``Peer Reflection" (PR) group
and the traditional group.
Each week in recitation, students in the PR group reflected in small teams on selected problems from the homework and 
discussed why solutions of some students
employed better problem solving strategies than others. The graduate and undergraduate teaching assistants (TAs) in the 
PR group recitations provided guidance and coaching to help students learn effective problem solving heuristics. 
In the recitations for the traditional group, 
students had the opportunity to ask the graduate TA questions about the homework before they took a weekly quiz. The traditional
group recitation quiz questions were similar to the homework questions selected for ``peer reflection" in the PR group recitations.
As one measure of the impact of this intervention, we investigated how likely students were to draw diagrams to help with problem solving.
On the final exam with only multiple-choice questions, the PR group drew diagrams on 
more problems than the traditional group, even when there was no external reward for doing so. 
Since there was no partial credit for drawing the diagrams on the scratch books, students did not draw diagrams simply to get 
credit for the effort shown and must value the use of diagrams for solving problems if they drew them. 
We also find that, regardless of whether the students belonged to the traditional or PR groups, those who drew more diagrams
for the multiple-choice questions 
outperformed those who did not draw them.

\end{abstract}

\section{Introduction}

\vspace{-.03in}
Students must learn effective problem solving strategies in order to develop expertise in physics. Specifically, they must be able
to solve problems beyond those that can be solved using a plug-and-chug approach.~\cite{fred1,fred2,fred3,fred4,larkin1,larkin2,maloney}
Research shows that converting a problem from the initial verbal representation to other suitable representations such as diagrammatic, tabular,
graphical or algebraic can make further analysis of the problem easier.~\cite{rep1,rep2,kaput} 
Similarly, using analogies or considering limiting cases are also useful strategies for 
solving problems.~\cite{gick1,gick4,intuition}
Many traditional courses do not explicitly teach students effective problem solving heuristics.
Rather, they may implicitly reward inferior problem solving strategies that many students engage in.
Instructors may implicitly assume that students appreciate the importance of initial qualitative analysis, planning, evaluation, and reflection 
phases of problem solving and that these phases are as important as the implementation phase.~\cite{schoenfeld1,schoenfeld2,schoenfeld3,schoenfeld4}
Consequently, they may not explicitly discuss and model these strategies while solving problems in class.
Recitation is usually taught by the teaching assistants (TAs) who present homework solutions
on the blackboard while students copy them in their notebooks.
Without guidance, most textbook problems do not help students
monitor their learning, reflect upon the problem solving process and pay attention to their knowledge structure.

Quantitative and conceptual problem solving both can enhance problem solving and reasoning skills, but only if
students engage in effective problem solving strategies rather than treating the task purely as a mathematical chore or 
guess-work~\cite{chi,mestre,hardiman,categorization,isomorph2}.
Without guidance, many introductory physics students do not perceive problem solving as an opportunity for learning to
interpret the concepts involved and to draw meaningful inferences from them.
Instead, they solve problems using superficial clues and cues, and apply concepts at random without concern for their applicability.
With explicit training, these same problem solving tasks can be turned into learning experiences that help students organize 
new knowledge coherently and hierarchically.
The abstract nature of the laws of physics and the chain of reasoning required to draw meaningful inferences make it even more important
to teach students effective problem solving strategies explicitly.

Reflection is an integral component of effective problem solving.~\cite{black1,black2}
While experts in a particular field
reflect and exploit problem solving as an opportunity for organizing and extending their knowledge, students often
need feedback and support to learn how to use problem solving as an opportunity for learning. 
There are diverse strategies that can be employed to help students reflect upon problem solving.
One approach that has been found to be useful is ``self-explanation" or explaining what one is learning explicitly to oneself.~\cite{self} 
Chi et al. found that, while reading science texts, students who 
constantly explained to themselves what they were reading and made an effort to connect the material read to their prior knowledge
performed better on problem solving on related topics given to them after the reading.~\cite{self} 
Inspired by the usefulness of self-explanation, Yerushalmi et al.
investigated how students may benefit from being explicitly asked to diagnose mistakes in their own quizzes with different levels of
scaffolding support.~\cite{edit1} 
They found that students benefited from diagnosing their own mistakes. The level of scaffolding
needed to identify the mistakes and correct them depended on the difficulty of the problems.

Another activity that may help students learn effective problem solving strategies while simultaneously learning physics content
is reflection with peers. In this approach, students
reflect not only on their own solution to problems, but reflect upon their peers' solutions as well.  
Integration of peer interaction (PI) with lectures has been popularized in the physics community by Mazur from Harvard 
University.~\cite{eric,eric2,eric3,eric4}
In Mazur's PI approach, the instructor poses conceptual problems in the form of multiple-choice questions to students 
periodically during the lecture.~\cite{eric,eric2,eric3,eric4} The focal point of the PI method is the discussion among students, which
is based on conceptual questions; the lecture component is limited and intended to supplement the self-directed learning.
The conceptual multiple choice questions give students an opportunity to think about the physics concepts and principles
covered in the lecture and discuss their answers and reasoning with peers. The instructor polls the class after peer interaction
to obtain the fraction of students with the correct answer.
On one hand, students learn about the level of understanding that is desired by the instructor by discussing with each other
the concrete questions posed. The feedback obtained by the instructor is also invaluable because the instructor learns about
the fraction of the class that has understood the concepts at the desired level.~\cite{eric}
This PI strategy keeps students alert during lectures and helps them monitor their learning, because not only do students
have to answer the questions, they must explain their answers to their peers.~\cite{eric} 
The method keeps students actively engaged in the learning process and
lets them take advantage of each others' strengths.  It helps both the low and high performing students,
because explaining and discussing concepts with peers helps students organize and solidify concepts in their minds.~\cite{eric}

Heller et al. have shown that group problem solving is especially valuable both for learning physics and for developing effective problem
solving strategies.~\cite{heller1,heller2} They have developed many
``context-rich'' problems that are close to everyday situations and are more challenging and stimulating than the
standard textbook problems. These problems require careful thought and the use of many
problem representations. Working with peers in heterogeneous groups with students with high, low and medium performance
is particularly beneficial for learning from the 
``context-rich" problems and students are typically assigned the rotating roles of manager, time keeper and skeptic by the 
instructor.~\cite{heller1,heller2}

Our prior research has shown that, even with minimal guidance from the instructors, students can benefit from peer interaction.~\cite{singhgroup}
In our study, those who worked with peers not only outperformed an equivalent group of students who worked alone on the same task, 
but collaboration with a peer led to ``co-construction" of knowledge in $29\%$ of the cases.~\cite{singhgroup} 
Co-construction of knowledge occurs when neither student who engaged in the peer collaboration was able to answer the
questions before the collaboration, but both were able to answer them after working with a peer on a post-test given individually to each person.

Here, we describe a study in which algebra-based introductory physics students in the Peer Reflection group (PR group) 
were provided guidance and support to reflect upon problem solving with peers and undergraduate and graduate teaching assistants 
in the recitation class. On the other hand, other recitation
classes were run in a traditional manner with the TA answering students' homework questions and then giving a quiz at the end
of each recitation class. Our assessment method was novel in that it involved counting the number of problems in which students drew diagrams
or did scratchworks on scratch books when there was no partial credit for these activities because the questions were in the multiple-choice format.
We find that the PR group drew more diagrams than the traditional group (statistically significant) even when there was no external reward for drawing 
them. 
The peer-reflection process which was sustained throughout the semester 
requires students to evaluate their solutions and those of their peers and involves high level of mental processing.~\cite{bloom}
The reflection process with peers can also help students monitor their learning.
We also find that there is a positive correlation between the number of diagrams drawn and the final exam performance. In particular,
students who drew diagrams for more problems performed 
better than others regardless of whether they belonged to the traditional group or the PR group. 

\section{Methodology}

The investigation involved an introductory algebra-based physics course mostly taken by students 
with interest in health related professions.  The course had 200 students and was broken into two sections both of which
met on Tuesdays and Thursdays and were taught by the same professor who had taught both sections of the course before. 
A class poll at the beginning of the course indicated that more than $80\%$ of the students had taken at least one physics course in 
high school, and perhaps more surprisingly, more than $90\%$ of the students had taken at least one calculus course (although the college 
physics course in which they were enrolled was an algebra-based course).

The daytime section taught during the day was the traditional group and had 107 students whereas the evening section called the
``Peer Reflection" group or PR group had 93 students. 
The lectures, all homework assignments, the midterm exams and the final exam were identical for the daytime and evening sections of the course.
Moreover, the instructor emphasized effective problem solving strategies, e.g., performing a conceptual
analysis of the problem and planning of the solution before implementing the plan and importance of evaluating the solution throughout the semester
in both the traditional and peer-reflection groups. 

Each week, students in both groups were supposed to turn in answers to the assigned homework problems (based upon the material covered in
the previous week) using an online homework system for some course credit. In addition, students in both groups were supposed to submit
a paper copy of the homework problems which had the details of the problem solving approach at the
end of the recitation class to the TA for some course credit. While the online homework solution was graded for correctness, the TA only 
graded the paper copies of the submitted homework for completeness on a three point scale (full score, half score or zero).

The weighting of each component of the course, e.g., midterm exams, final exam, class participation, homework and the scores allocated for the recitation 
were the same for both classes. Also, as noted earlier, all components of the course were identical for both groups except
the recitations which were conducted very differently for the PR and traditional groups.  
Although the total course weighting assigned to the recitations was the same for both groups (since all the other components of the course had the 
same weighting for both groups), the scoring of the recitations was different for the two groups. 
Students were given credit for attending recitation in both groups. 
Attendance was taken in the recitations using clickers for both the traditional group and the PR group. 
The traditional group recitations were traditional in which the TA would solve selected assigned homework problems on the blackboard
and field questions from students about their homework before assigning a quiz in the last 20 minutes of the recitation class. 
The recitation quiz problems given to the traditional group were similar to the homework problems selected for ``peer 
reflection" in the PR group recitations (but the quiz problems were not identical to the homework problems to discourage students in the 
traditional group from memorizing the answers to homework in preparation for the quiz). 
Students in the PR group reflected on three homework problems in each recitation class but no recitation quiz was given to the 
students in this group at the end of the recitation classes, unlike the traditional group, primarily due to the time constraints.
The recitation scores for the PR group students were assigned based mostly on the recitation attendance except students obtained bonus points for
helping select the ``best" student solution as described below.
Since the recitation scoring was done differently for the traditional and PR groups, the two groups
were curved separately so that the top $50\%$ of the students in each group obtained A and B grades in view of the departmental policy.

As noted earlier, both recitation sections for the evening section (93 students total) together formed the 
PR group. The PR group intervention was based upon a field-tested cognitive 
apprenticeship model~\cite{cog1,cog2} of learning involving modeling, coaching, and fading to help students learn effective 
problem solving heuristics. In this approach, ``modeling" means that the TA demonstrates and exemplifies the effective problem solving 
skills that the students should learn. ``Coaching" means providing students opportunity to practice problem solving skills
with appropriate guidance so that they learn the desired skills.  
``Fading" means decreasing the support and feedback gradually with a focus on helping students develop self-reliance.

The specific strategy used by the students in the PR group involved
reflection upon problem solving with their peers in the recitations, while the TA and the undergraduate teaching assistants (UTAs)
exemplified the effective problem solving heuristics. The UTAs were chosen from those undergraduate students who had earned an $A^+$ grade in
an equivalent introductory physics course previously. The UTAs had to attend all the lectures in the semester in which they were UTAs for
a course and they communicated with the TA each week (and periodically with the course instructor) to determine the plan for the recitations. 
We note that, for effective implementation of the PR method, two UTAs were present in each recitation class along with the TA. These 
UTAs helped the TA in demonstrating and helping students to learn effective problem solving heuristics. 

In our intervention, each of the three recitation sections in the traditional group had about 35-37 students. 
The two recitations for the PR group had more than 40 students each (since the PR group was the evening section of the course, it was logistically not possible to break this group into three recitations). At the 
beginning of each PR recitation, students were asked to form nine teams of three to six
students chosen at random by the TA (these teams were generated by a computer program each week). 
The TA projected the names of the team members on the screen so that they could sit together at the beginning 
of each recitation class. Three homework questions were chosen for a particular recitation.  
The recitations for the two sections were coordinated by the TAs so that the recitation quiz problems given to the traditional group were based upon 
the homework problems selected for ``peer reflection" in the PR group recitations. Each of the three ``competitions" was carefully timed to take 
approximately 15 minutes, in order for the entire exercise to fit into the allotted fifty-minute time slot.

After each question was announced to the class, each of the nine teams were given three minutes to identify the ``best" solution by comparing and 
discussing among the group members.  If a group had difficulty coming up with a ``winner", the TA/UTA would intervene and facilitate the process.  
The winning students were asked to come to the front of the room, where they were assembled into three second-round groups.  
The process was repeated, producing three finalists. These students handed in their homework solutions to the TAs, after which the 
TA/UTA evaluation process began. A qualitative sketch of the team structures at various stages of the competition is shown in Figure 1.

The three finalists' solutions were projected one at a time on a screen using a web cam and computer projector. Each of the three panelists (the TA
and two UTAs) gave 
their critique of the solutions, citing what each of the finalists had done well and what could be done to further enhance the problem solving 
methodology in each case.  In essence, the TA and UTAs were ``judges" similar to the judges in the television show ``American Idol" and gave their 
``critique" of each finalist's problem solving performance. After each solution had been critiqued by each of the panelists, the students, using 
the clickers, voted on the ``best" solution. The TA and UTAs did not participate in the voting process. 

In order to encourage each team in the PR group to select the student with the most effective problem solving strategy as the winner for each 
problem, all students from the teams whose member advanced to the final round to ``win" the ``competition" were given course credit (bonus points). 
In particular, each of these team members (consolation prize winners) earned one third of the course credit given to the student whose solution was declared to be the ``winner".
This reward system made the discussions very lively and the teams generally 
made good effort to advance the most effective solution to the next stage. Figure 1 shows one possible team configuration at various stages of PR 
activities when there are 27 students in the recitation class initially. Due to lack of space, each of the initial teams (round 1) 
in Figure 1 is shown with 3 members whereas in reality this round on average consisted of five team members.  In each team, the student with the dark border in Figure 1 is the
``winner of that round" and advances to the 
next stage. All the students who participated at any stage in helping select the ``winner" (those shown in gray) were the consolation
prize winners and obtained one third of the course credit that was awarded to the winner for that problem.

While we video-taped a portion of the recitation class discussions when students reflected with peers, a good account of the effectiveness and 
intensity of the team discussions came from the TA and UTAs who generally walked around from team to team listening to the discussions but not 
interrupting the team members involved in the discussions unless facilitation was necessary for breaking a gridlock. 
The course credit and the opportunity to have the finalists' solutions voted on by the whole class encouraged students to argue passionately about 
the aspects of their solutions that displayed effective problem solving strategies. Students were constantly arguing about why drawing a diagram, 
explicitly thinking about the knowns and target variables, and explicitly justifying the physics principles that would be useful before writing down the 
equations are effective problem solving strategies.

Furthermore, the ``American Idol" style recitation allowed the TAs to discuss and convey to students in much more detail what solution styles were 
preferred and why.  Students were often shown what kinds of solutions were easier to read and understand, and which were more amenable to 
error-checking.  Great emphasis was placed on consistent use of notation, setting up problems through the use of symbols to define physical 
quantities, and the importance of clear diagrams in constructing solutions.

At the end of the semester, all of the students were given a final exam consisting of 40 multiple choice questions, 20 of which were primarily 
conceptual in nature and 20 of which were primarily quantitative (students had to solve a numerical or symbolic problem for a target quantity). 
Although the final exam was all multiple-choice, a novel assessment method was used. While students knew that the only thing that counted
for their grade was whether they chose the correct option for each multiple-choice question, each student was given an exam notebook which
he/she could use for scratchworks. 
We hypothesized that even if the final exam questions were in the multiple-choice format, students who value effective problem
solving strategies will take the time to draw more diagrams and do more scratchworks even if there was no course credit for such activities. 
With the assumption that the students will write on the exam booklet and write down relevant concepts only if they think it is helpful for
problem solving, multiple-choice exam can be a novel tool for assessment. It allowed us to observe students' problem solving strategies in a 
more ``native" form closer to what they really think is helpful for problem solving 
instead of what the professor wants them to write down or filling the page when a free-response
question is assigned with irrelevant equations and concepts with the hope of getting partial credit for the effort.

We decided to divide the students' work in the notebooks and exam-books into two categories: diagrams and scratchworks. 
The scratchworks included everything written apart from the diagrams such as equations, sentences, and texts.
Both authors of this paper agreed on how to differentiate between diagrams and scratchworks.
Instead of using subjectivity in deciding how ``good" the diagrams or scratchworks for each student for each of the 40 questions were,
we only counted the number of problems with diagrams drawn and scratchworks done by each student. For example, if a student drew diagrams
for 7 questions out of 40 questions and did scratchworks for 10 questions out of 40 questions, we counted it as 7 diagrams and 10 scratchworks. 

\section{Motivation and Goal}

We hypothesized that the PR intervention in the recitation may be beneficial in helping students learn effective problem solving strategies 
because students solved the homework problems, discussed them with their peers to determine the top three student solutions and then were
provided expert feedback from the UTAs and TA about those solutions. Chi et al.~\cite{glaser} have theorized that 
students often fail to learn from the ``expert" solutions even when they realize that their own approaches are deficient 
because they may not take the time to reflect upon how the expert model is different from their own and how they can repair their
own knowledge structure and improve their problem solving strategies. 
Chi et al. therefore emphasize that simply realizing that the problem solving 
approach of the instructor is superior than theirs may not be enough to help students learn effective strategies. They noted that the
feedback and support by instructors
to help students understand why the expert solution is better after the students realize that they are better
is a critical component of learning effective approaches to problem solving.~\cite{glaser}
In the approach adopted in the PR group recitations,
the UTAs and TA provided valuable ``expert" feedback to help students make these connections right after the students had thought about 
effective approaches to problem solving themselves. In fact, since high performing UTAs are only a year ahead of the students, their
feedback may be at the right level and even more valuable to the students.~\cite{eric}
Such reasoning was our motivation for exploring the effect of the PR intervention.

Our goal was to examine both inter-group effects and group-independent effects.
Inter-group effects refer to the investigation of the differences between the traditional group and the PR group.
For example, we investigated whether there was a statistical difference in the average number of problems 
for which students drew diagrams and wrote scratchworks in the PR group and the traditional group. 
We also examined group-independent effects, findings that hold for students in both the traditional group and the PR group.
One issue we examined was whether students who drew more diagrams, despite knowing that there was no partial credit for these tasks, were more
likely to perform better in the final exam. 
We also investigated the correlation between the number of problems with diagrams or scratchworks and the final
exam performance when quantitative and conceptual questions were considered separately.
We also explored whether students were more likely to draw diagrams on quantitative or conceptual questions on the final exam.

\section{Results}

Although no pretest was given to students, there is some evidence that, over the years, the evening section
of the course is somewhat weaker and does not perform as well overall as the daytime section of the course historically. 
The difference between the daytime and evening sections of the 
course could partly be due to the fact that some students in the evening section work full-time and take classes simultaneously.
For example, the same professor had also taught both sections of the course one year before the peer reflection activities were
introduced in evening recitations and thus all recitations for both sections of the course were taught traditionally that year. 
Thus, we first compare the averages of the daytime and evening sections before and after the peer reflection activities were instituted in
the evening recitation classes.  Table 1 compares the difference in the averages between the 
daytime and evening classes the year prior to the introduction of peer reflection (Fall 2006) 
and the year in which peer reflection was implemented in the evening recitation classes (Fall 2007). 
In Table 1, the p-values given are the results of t-tests performed between the daytime and evening classes.~\cite{stat}  
Statistically significant difference (at the level of $p=0.05$) between groups only exists between the average midterm scores for the year 
in which peer reflection was implemented. 
The evening section scored lower on average than the daytime section on the final exam but the difference is not statistically significant
(p=0.112 for 2006 and p=0.875 for 2007), as indicated in Table 1. 
We note that while the midterm questions differed from year to year (since the midterms were returned to students and there was a possibility
that the students would share them), the final exam, which was not returned to students,
was almost the same both years (except the instructor had changed a couple of questions on the final exam from 2006 to 2007).
We also note that the final exam scores are lower than the midterm scores because there was no partial credit for answering
multiple-choice questions on the final exam.

The final exam which was comprehensive had 40 multiple-choice questions, half of which were quantitative and half were conceptual.
There was no partial credit given for drawing the diagrams or doing the scratchworks. One issue we investigated is whether the students considered the 
diagrams or the scratchworks to be beneficial and used them while solving problems, even though students knew that no partial credit was given for showing work.
As noted earlier, our assessment method involved counting the number of problems with diagrams and scratchworks. We counted any 
comprehensible work done on the exam notebook other than a diagram as a scratchwork. In this sense, quantifying the amount of scratchwork does not 
distinguish between a short and a long scratchwork for a given question. If a student wrote anything
other than a diagram, e.g., equations, the known variables and target quantities, an attempt to solve for unknown etc., 
it was considered scratchwork for that problem. Similarly, there was a diversity in the quality of diagrams the students
drew for the same problem. Some students drew elaborate diagrams which were well labeled while others drew rough sketches. Regardless of
the quality of the diagrams, any problem in which a diagram was drawn was counted.

We find that the PR group on average drew more diagrams than the traditional group.
Table 2 compares the average number of problems with diagrams or scratchworks by the traditional group and the PR group on the final exam. 
It shows that the PR group had more problems with diagrams than the traditional group (statistically significant)~\cite{stat}. 
In particular, the traditional group averaged 7.0 problems with diagrams per student for the whole exam (40 problems), 4.3 problems
with diagrams for the quantitative questions per student and 2.7 problems with diagrams 
for the conceptual questions per student. On the other hand, the PR group averaged 8.6 problems with diagrams per student, 5.1 
problems with diagrams for the quantitative questions per student and 3.5 diagrams for the conceptual questions per student. 
Figures 2 and 3 display the histograms of the total number of problems with diagrams on the final exam (with only multiple-choice questions)
vs. the percentage of students for the traditional and PR groups respectively. Figures 2 shows that some students in 
the PR group drew many more diagrams than the average number of diagrams drawn on the exam.
The histograms of the total number of quantitative or conceptual questions with diagrams vs. the percentage of students for
the traditional and PR groups can be found here.\cite{epaps}

We also find that students drew more diagrams for the quantitative questions than the conceptual questions.
Tables 2 shows that, regardless of whether they belonged to the traditional group or the PR group, students were more likely to draw diagrams for 
the quantitative questions than for the conceptual questions. 
The comparison of the number of problems with diagrams for quantitative vs. conceptual problems 
for each of the traditional (N=107) and PR (N=93) groups displayed in Table 2 gives statistically significant difference at the level of $p=0.000$. 
It was not clear {\it a priori} that students will draw more diagrams for the 
quantitative questions than for the conceptual questions and we hypothesize that this trend may change depending on the expertise of the 
individuals and the type of questions asked.
Table 2 also shows that, while there was no statistical difference between the two groups in terms of the total number of problems with 
scratchworks performed, the students were far more likely to do scratchworks for the quantitative questions than for the conceptual questions. 
Figures 4 and 5 display the histograms of the total number of problems with scratchworks on the final exam 
vs. the percentage of students for the traditional and PR groups respectively. 
The histograms of the total number of quantitative or conceptual questions with scratchworks vs. the percentage of students for
the traditional and PR groups can be found here.\cite{epaps} 
We note that the sample sizes are approximately equal and not too small,
so the t-test will not be much affected even if the distributions
are skewed.~\cite{hopkins}
The comparison of the number of problems with scratchworks for quantitative vs. conceptual problems 
for each of the traditional and PR groups displayed in Table 2 gives statistically significant difference at the level of $p=0.000$. 
This difference between the quantitative and the conceptual questions
makes sense since problems which are primarily quantitative require calculations to arrive at an answer.

We also find a positive correlation between the final exam score and the number of problems with diagrams.
Table 3 investigates whether the final exam score is correlated with the number of problems with diagrams or scratchworks
for the traditional group and the PR group separately (R is the correlation coefficient). The null hypothesis in each case is 
that there is no correlation between
the final exam score and the variable considered, e.g., the total number of problems with diagrams drawn.~\cite{stat} 
Table 3 shows that, for both the traditional group and the PR group, the students who had more problems with diagrams and scratchworks were
statistically (significantly) more likely to perform well
on the final exam.~\cite{stat} This correlation holds regardless of whether we look at all questions or the quantitative questions only (labeled diagram (Q) 
and scratch(Q) for the quantitative diagrams and scratchworks respectively) or conceptual questions only (labeled diagram(C) and scratch(C)
for the conceptual diagrams and scratch work respectively).
There is no prior research related to physics learning at any level that we know of
that shows a positive correlation between the exam performance and the number of problems with diagrams drawn
when answering multiple-choice questions where there is no partial credit for drawing diagrams. 

It is evident from Table 3 that for both conceptual and quantitative questions, diagrams are positively correlated with final exam performance. 
In particular, we investigated the correlations between the number of problems with diagrams drawn (or the amount of scratchwork) and the 
final exam scores 
on the quantitative questions alone and the conceptual questions alone. Table 3 shows that within each group (traditional or PR), the correlation 
between the number of diagrams drawn and the final exam score is virtually 
identical for the quantitative and conceptual questions (R = 0.19 for quantitative vs. R=0.20 for conceptual in the traditional group, R=0.36 for 
quantitative vs. R=0.36 for conceptual in the peer reflection group).  

We find that the diagrams drawn by the PR group explain more of the final exam performance.
In particular, the comparison of the traditional group and the PR group in Table 3 shows that for each case shown in the different rows, the correlation between
the number of diagrams or amount of scratchwork and the final exam score is stronger for the PR group than for the traditional group.
For example, the correlation coefficient $R$ for the number of problems with diagrams drawn vs. the final exam score is higher for 
the PR group compared to the traditional group (R=0.40 vs. R=0.24).
The PR group also showed a stronger correlation than the traditional group even when the quantitative and conceptual questions were
considered separately for the correlation between the number of problems with diagrams drawn and the final exam scores.
Similarly, the correlation coefficient for the number of scratchworks vs. the final exam score is higher for the PR group compared to the 
traditional group (R=0.53 vs. R=0.39). 

We also find that scratchworks explain more of the performance for quantitative questions than conceptual questions.
In particular, Table 3 also shows that, 
within each group (traditional and PR), the correlation between the amount of scratchwork and the final exam score was
stronger for the quantitative questions than for the conceptual questions. 
While correlation does not imply causality, the stronger correlation may be due to the fact that
students do not necessarily have to perform algebraic manipulations for conceptual questions but it may be a 
pre-requisite for identifying the correct answer for a quantitative question.
Further examination of the quantitative-only correlations between the scratchwork and the final exam scores
(R = 0.42 for the traditional group, R = 0.59 for the  PR group) 
shows that the correlations are stronger for the PR group than the traditional group.  

\section{Discussion and Summary}

In this study, the recitation classes for an introductory physics course primarily for the health-science majors were broken into a traditional
group and a ``peer reflection" or PR group.
We investigated whether students in the PR group use better problem-solving strategies such as drawing diagrams than
students in the traditional group and also whether there are
differences in the performance of the two groups. We also explored whether students who perform well are the ones who are more likely
to draw diagrams or write scratchworks even when there is no partial credit for these activities.

In the PR group recitation classes, students reflected about their problem solving in the homework with peers each week.
Appropriate guidance and support provided opportunities for learning effective problem solving heuristics to the students in the PR group.
In particular, students in the PR group reflected in small teams on selected problems from the homework and discussed why 
solutions of some students employed better problem solving strategies than others. 
The TA and UTAs in the PR group recitations demonstrated effective approaches to problem solving and coached students so that they learn 
those skills. 
Each small team in the PR group discussed which student's homework solutions employed the most effective problem solving heuristics and
selected a ``winner". Then, three teams combined into a larger team and repeated the process of determining
the ``winning" solution. Typically, once three ``finalists" were identified in this manner, the TA and UTAs put each 
finalist's solution on a projector and discussed what they perceived to be good problem-solving strategies used in each solution
and what can be improved. Finally, each student used clickers to vote on the ``best" overall solution with regard to the problem 
solving strategies used. There was a reward system related to course credit
that encouraged students to be involved in selecting the solution with the best problem solving strategy in each round.
Students in the traditional group had traditional recitation classes in which
they asked the TA questions about the homework before taking a quiz at the end of each recitation class. 
Each problem selected for ``peer reflection" was adapted into a quiz problem for the traditional group. 

The assessment of the effectiveness of the intervention was novel. The final exam had 40 multiple-choice questions, half of which
were quantitative and half were conceptual. 
For the multiple-choice questions, students technically do not have to show their work
to arrive at the answer. However, students may use effective approaches to problem solving such as drawing
a diagram or writing down their plan if they believe it may help them answer a question correctly.
Although students knew that there was no partial credit for drawing diagrams or writing scratchworks,
we compared the average number of problems with diagrams or scratchworks in the traditional and experimental groups.
Our hypothesis was that students who value effective problem solving strategies will have more problems in which diagrams are
drawn or scratchworks are written despite the fact that there is no partial credit for these activities. 
In fact, the fact that there was no partial credit for the diagrams or the scratchworks helped
eliminate the possibility of students drawing the diagrams or writing scratchworks for the sole purpose of getting partial credit for the effort 
displayed (even if it is meaningless from the perspective of relevant physics content).

We note that to help understand the statistical differences between the PR and traditional sections, 
we chose to quantify the number of diagrams drawn in 
the multiple-choice final examination.  It should be stressed that the PR recitations emphasized a wide variety of preferred problem-solving 
strategies (not just drawing diagrams).  Diagrams (and scratchworks) were chosen simply because they were more straightforward to quantify than 
other strategies.  We hypothesize that the use of diagrams represents a good ``marker" for overall improvement in problem solving strategies. 

Our findings can be broadly classified into inter-group and group-independent categories.
The inter-group findings that show the difference between the traditional group and PR group can be summarized as follows:
\begin{itemize}
\item On the multiple-choice final exam where there was no partial credit for drawing diagrams, the PR group drew diagrams in more problems
(statistically significant) than the traditional group. 
\item The diagrams drawn by the PR group explain more of the final exam performance than those by the traditional group.
\end{itemize}

Findings that are independent of group (which are true even when the traditional group and PR group are not separated and all students
are considered together) can be summarized as follows:
\begin{itemize}
\item There is a statistically significant positive correlation between how often students wrote scratchworks or drew diagrams and how well they 
performed on the final exam regardless of whether they were in the traditional group or the PR group. 
In particular, those who performed well in the multiple-choice final exam
(in which there was no partial credit for showing work) were much more likely to draw diagrams than the other students. 
While one may assume that high-performing students will draw more diagrams even when there is no partial credit
for it, no prior research that we know of has explicitly demonstrated a correlation between the number of ``genuinely drawn"
diagrams and student performance at any level of physics instruction.
\item The correlations between the number of problems with diagrams drawn and the final exam scores 
on the quantitative questions alone and the conceptual questions alone are comparable and positive. 
\item Students in both groups were more likely to draw diagrams or write scratchworks for quantitative problems than for the conceptual questions.
While more scratchworks are expected on quantitative problems, it is not clear {\it a priori} that more diagrams will be drawn for the quantitative
problems than for the conceptual questions. We hypothesize that this trend may depend upon the expertise of the individuals, explicit training in 
effective problem solving strategies and the difficulty of the problems.
\end{itemize}

We note that the students in the traditional group were given weekly recitation quizzes in the last 20 minutes of the recitation class
based upon that week's homework. There were no recitation quizzes in the PR group due to the time constraints. 
It is sometimes argued by faculty members that the recitation quizzes are essential to keep students engaged in the learning process during the 
recitations. However, this study shows that the PR group was not adversely affected by not having the weekly quizzes that the traditional group had 
and instead having the peer reflection activities. The mental engagement of students in the PR group throughout the recitation class
may have more than compensated for the lack of quizzes. The students in the PR group were evaluating their peer's work along with their
own which requires high level of mental processing.~\cite{bloom} They were comparing problem solving strategies such as how to do a conceptual
analysis and planning of the solution, why drawing two separate diagrams may be better in certain cases (e.g., before and after a collision) 
than combining the information into one diagram, how to define and use symbols consistently etc. In addition, after the active engagement with peers, 
students got an opportunity to learn from the TA and UTAs about their critique of each ``winning" solution highlighting the strengths and weaknesses. 
We also note that the PR recitations do not require any additional class time or effort on the part of the instructor. 

According to Chi~\cite{glaser}, students are likely to improve their approaches to problem solving and learn meaningfully from an intervention if both
of the following happen: I) students compare two artifacts, e.g., the expert solution and their own solution and realize their omissions,
 and II) they receive guidance to understand why the expert solution is better and how they can improve upon their
own approaches. The PR approach uses such a two tier approach in which students first identify that other student's solution may be
better than their own and then are guided by the UTAs/TA to reflect upon the various aspects of the ``winning" solutions. 

\vspace*{-.12in}
\section{Acknowledgments}
\vspace*{-0.2in}

We are very grateful to Dr. Louis Pingel from the School of Education and Dr. Allan R. Sampson and Huan Xu from the Department of Statistics at 
the University of Pittsburgh
for their help in data analysis. We thank F. Reif, R. Glaser, R. P. Devaty and J. Levy for helpful discussions.
We thank J. Levy for his enthusiastic support of helping us implement the ``peer reflection" activities in his class.
We thank the National Science Foundation for financial support.

\pagebreak

\begin{center}
\epsfig{file=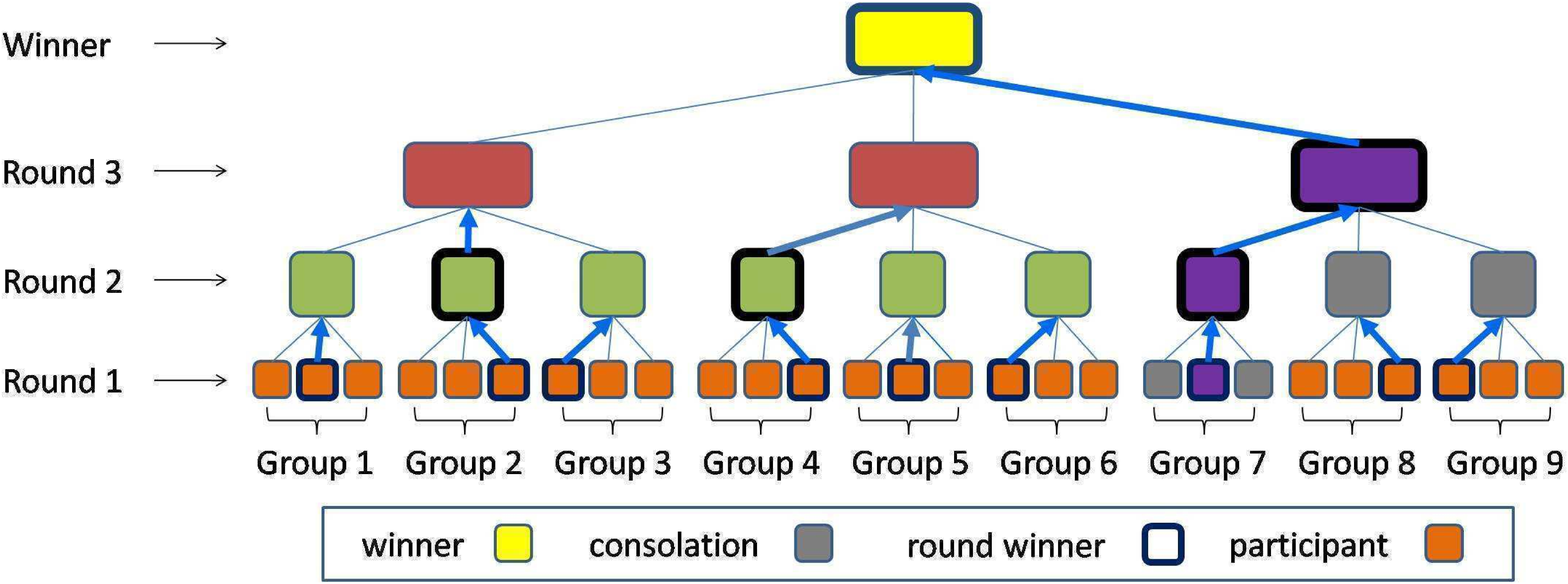,height=2.5in}
\end{center}
Figure 1: Illustration of the team structure at the three stages of peer-reflection activities. Before the students voted in the third round 
using the clickers, the TA and UTAs critiqued each of the three solutions at the final stage. Due to lack of space, only 3 team members per team are 
shown in round 1 but there were on average five members in each group in this round (as generated by a computer program each week). The consolation
prize winners in gray obtained 1/3rd of the course credit awarded to the winner.

\pagebreak

\begin{center}
\epsfig{file=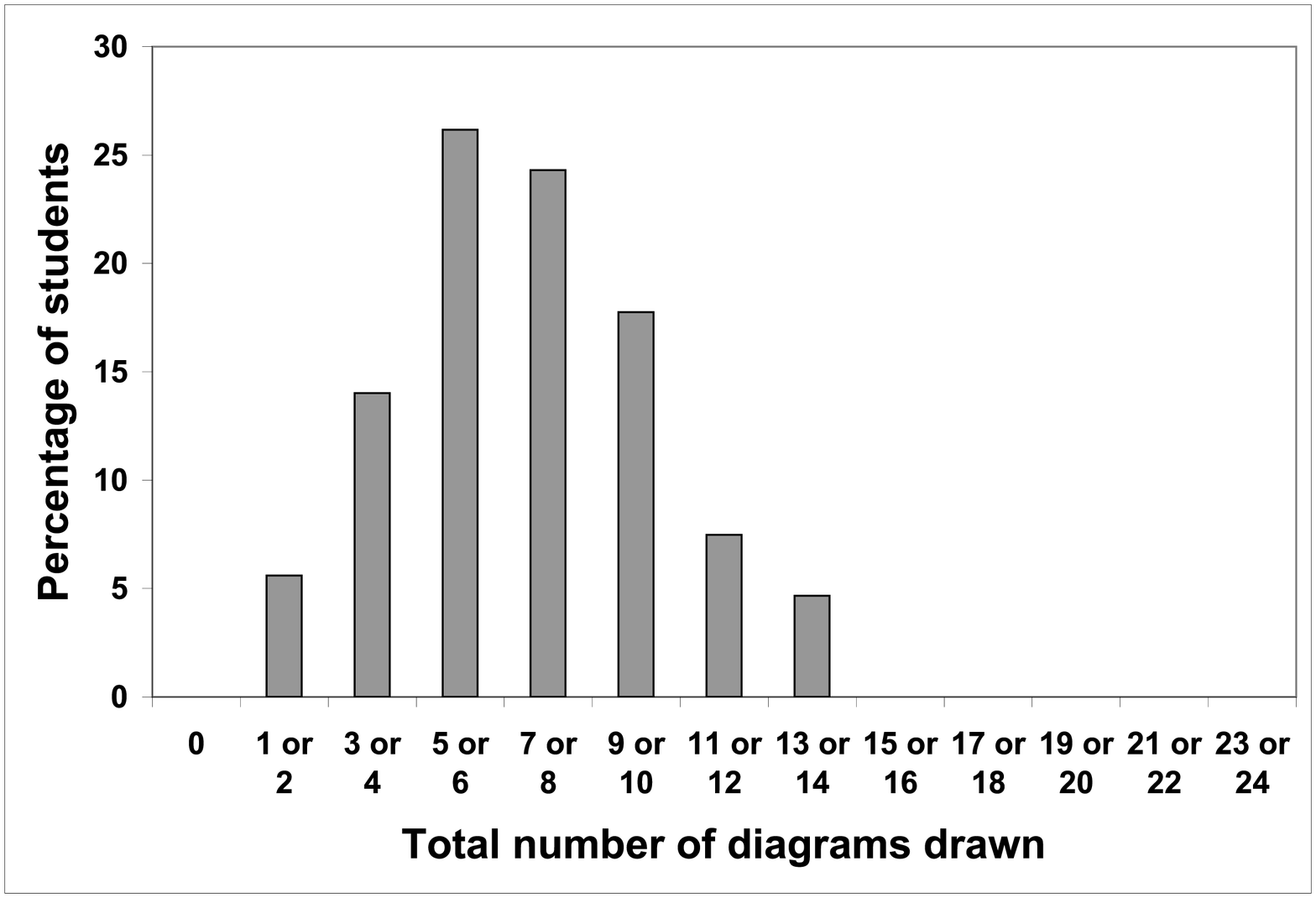,height=3.0in}
\end{center}
Figure 2: Histogram of the total number of questions with diagrams drawn on the final exam (consisting solely of multiple-choice problems) vs. percentage of students for
the traditional group.

\pagebreak

\begin{center}
\epsfig{file=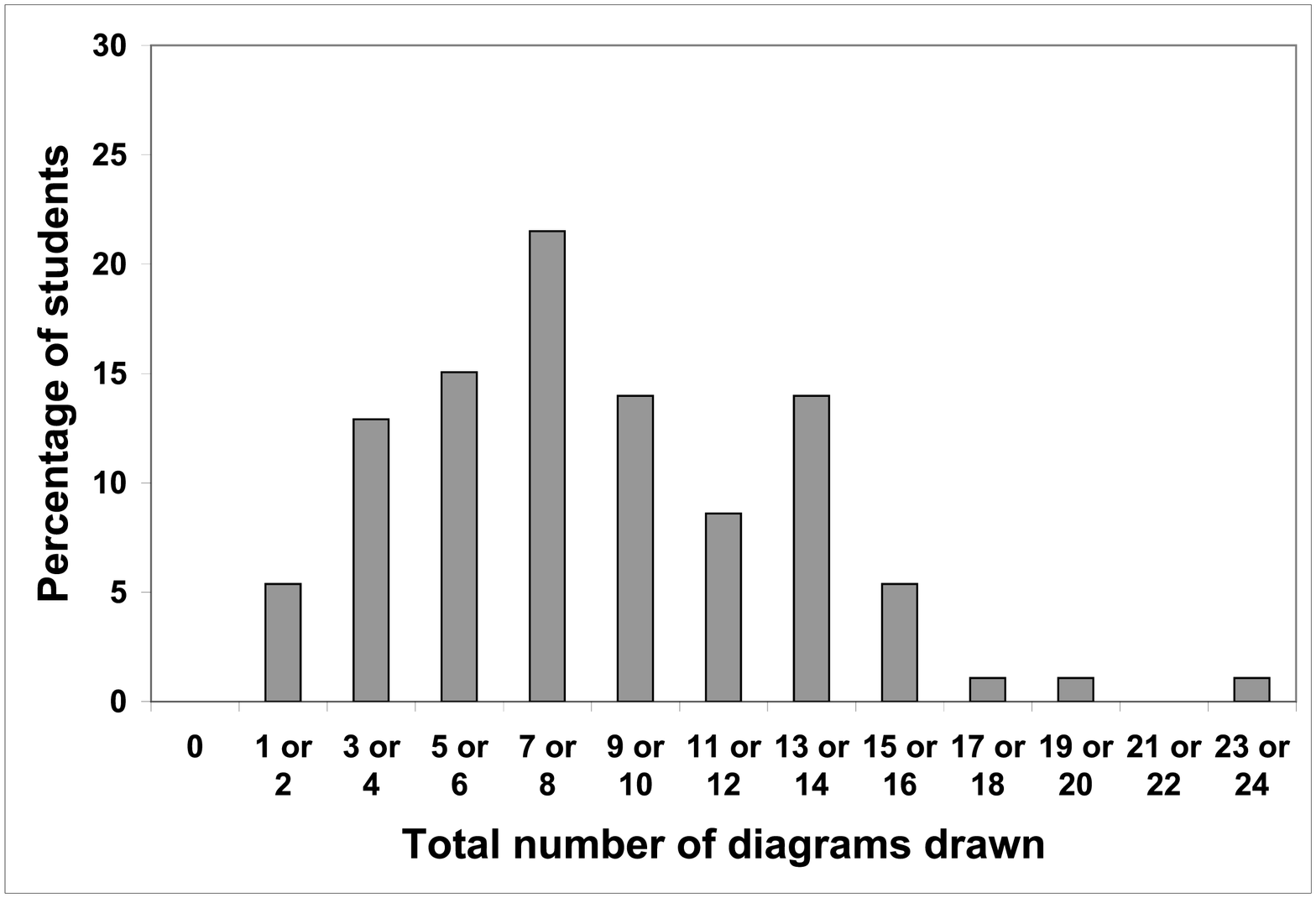,height=3.0in}
\end{center}
Figure 3: Histogram of the total number of questions with diagrams drawn on the final exam (consisting solely of multiple-choice problems) vs. percentage of students for
the PR group.

\pagebreak

\begin{center}
\epsfig{file=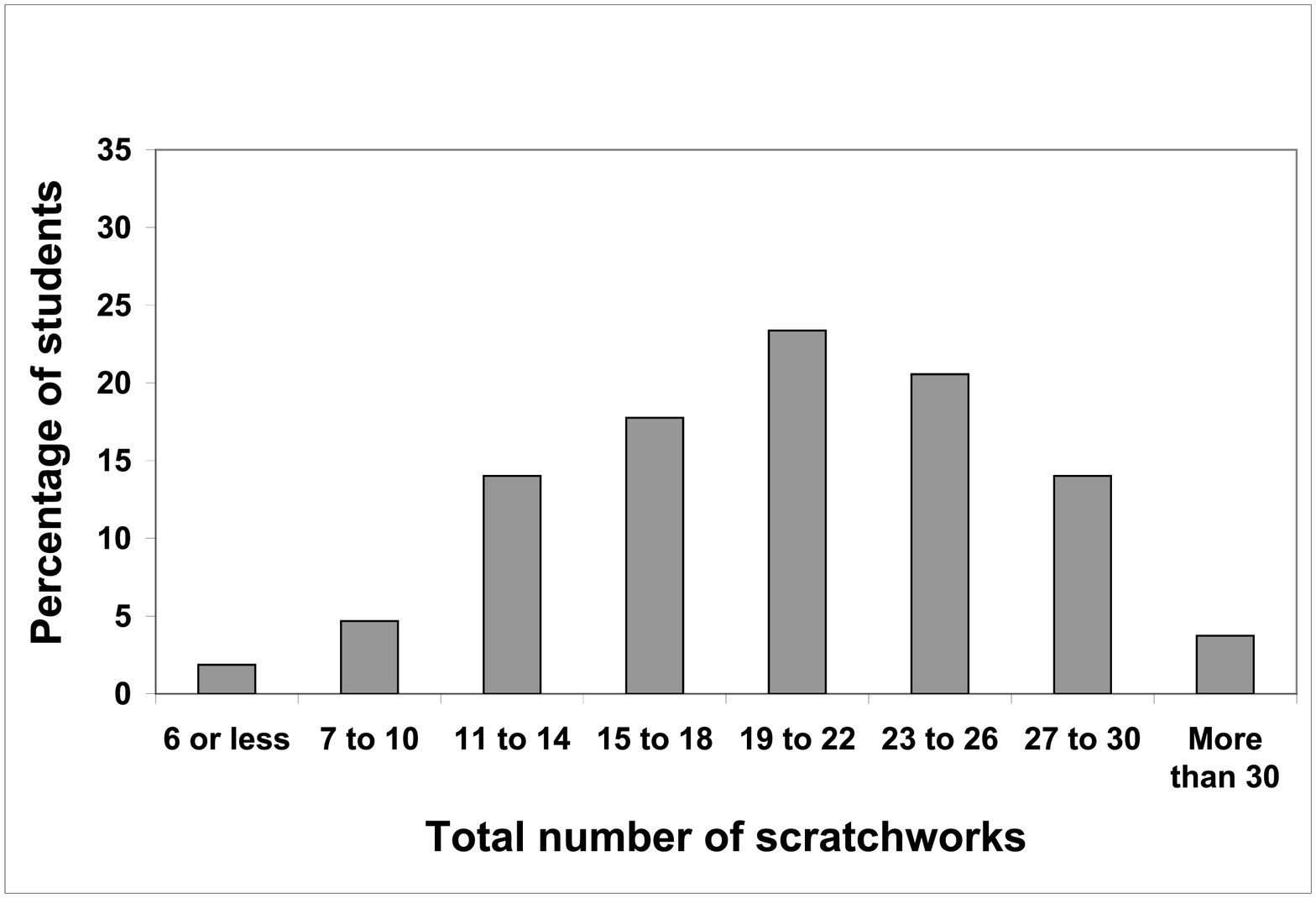,height=3.0in}
\end{center}
Figure 4: Histogram of the total number of questions with scratchworks on the final exam (consisting solely of multiple-choice problems) vs. percentage of students for the traditional group.

\pagebreak

\begin{center}
\epsfig{file=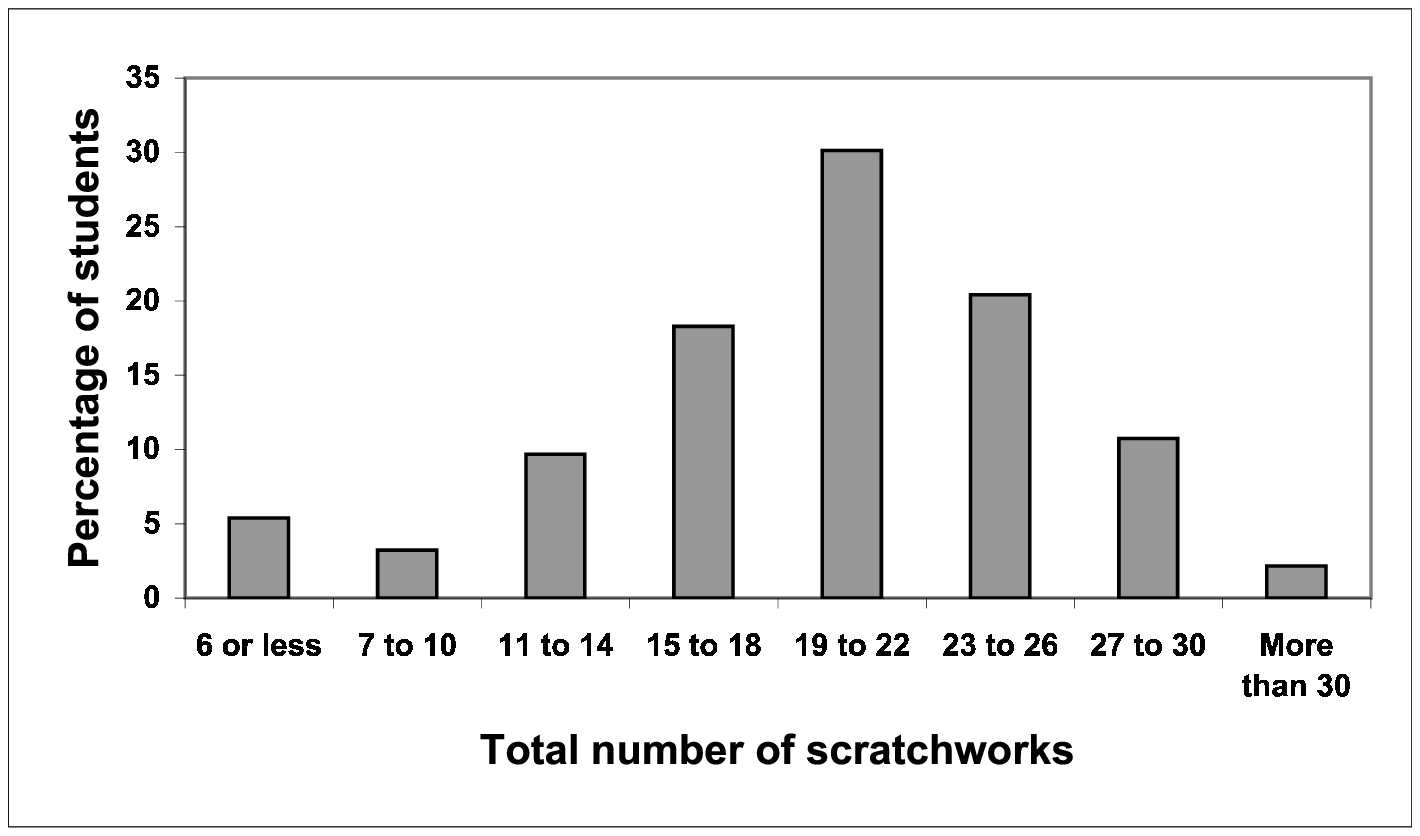,height=2.5in}
\end{center}
Figure 5: Histogram of the total number of questions with scratchworks on the final exam (consisting solely of multiple-choice problems) vs. percentage of students for the PR group.

\pagebreak

\begin{table}[h]
\centering
\begin{tabular}[t]{|c|c|c|c|}
\hline
Daytime vs. Evening classes& Daytime& Evening& p-value  \\[0.5 ex]
& means \%& means\% & \\[0.5 ex]
\hline
2006: midterm exams& 72.0&65.8 (non-PR)&0.101\\[0.5 ex]
\hline
2006: final exams& 55.7&52.7 (non-PR)&0.112\\[0.5 ex]
\hline
2007: midterm exams& 78.8&74.3 (PR)&0.004\\[0.5 ex]
\hline
2007: final exams& 58.1&57.7 (PR)&0.875\\[0.5 ex]
\hline
\end{tabular}
\vspace{0.1in}
\caption{Means and p-values for comparisons of the daytime and evening classes
during the year before peer reflection was introduced (Fall 2006) and during the year in which 
it was introduced (Fall 2007). The following were the number of students in each group: Fall 2006 daytime N=124, evening N=100, Fall 2007 daytime N=107,
evening N=93.}
\label{junk2}
\end{table}

\pagebreak

\begin{table}[h]
\centering
\begin{tabular}[t]{|c|c|c|c|c|}
\hline
& Question type & Traditional group & PR group& p-value  \\[0.5 ex]
& & per & per & between  \\[0.5 ex]
& & student & student& groups  \\[0.5 ex]
\hline
& All questions & 7.0& 8.6& 0.003  \\[0.5 ex]
& (40 total)& & &   \\[0.5 ex]\cline{2-5}
Number of & Quantitative & 4.3& 5.1& 0.006  \\[0.5 ex]
problems with & (20 total)& & &  \\[0.5 ex]\cline{2-5}
diagrams& Conceptual& 2.7 & 3.5& 0.016  \\[0.5 ex]
&  (20 total) & & &  \\[0.5 ex]\cline{2-5}
\hline
& All questions & 20.2& 19.6& 0.496  \\[0.5 ex]
& (40 total)& & &   \\[0.5 ex]\cline{2-5}
Number of & Quantitative & 16.0 & 15.6 &0.401    \\[0.5 ex]
problems with & (20 total)& & &   \\[0.5 ex]\cline{2-5}
scratchworks& Conceptual& 4.2 & 4.0 &0.751  \\[0.5 ex]
&  (20 total) & & &   \\[0.5 ex]\cline{2-5}
\hline
\hline
\end{tabular}
\vspace{0.1in}
\caption{Comparison of the average number of problems per student with diagrams and scratchworks by the traditional group (N=107) and the PR group 
(N=93) in the final exam. The PR group has significantly more problems with diagrams than the traditional group. The average number of problems
with scratchworks per student in the two groups is not significantly different. 
There are more quantitative problems with diagrams drawn and scratchworks written than conceptual problems (at the level of $p=0.000$).}
\label{junk2}
\end{table}

\pagebreak

\hspace*{-1.5in}
\begin{table}[h]
\centering
\begin{tabular}[t]{|c|c|c|c|c|}
\hline
& \multicolumn{2}{|c|}{Traditional: Final }  & \multicolumn{2}{|c|}{ PR: final}  \\[0.5 ex]\cline{2-5}
& R& p-value& R& p-value \\[0.5 ex]
\hline
Diagram & 0.24 & 0.014 & 0.40 & 0.000 \\[0.5 ex]
\hline
Diagram(Q) & 0.19 & 0.046 & 0.36 & 0.000 \\[0.5 ex]
\hline
Diagram(C) & 0.20 & 0.042 & 0.36 & 0.000 \\[0.5 ex]
\hline
Scratch & 0.39 & 0.000 & 0.53 & 0.000 \\[0.5 ex]
\hline
Scratch(Q) & 0.42 & 0.000 & 0.59 & 0.000 \\[0.5 ex]
\hline
Scratch(C) & 0.28 & 0.004 & 0.32 & 0.002 \\[0.5 ex]
\hline
\end{tabular}
\vspace{0.1in}
\caption{Correlation between the final exam scores and different variables such as the number of problems with diagrams or scratchworks for
each of the traditional group (N=107) and the PR group (N=97). 
The students who performed well on the final exam in both the traditional group and the PR group
had significantly more problems with diagrams and scratchworks than the students who performed poorly on the final exam.
}
\label{junk2}
\end{table}

\end{document}